\documentclass[journal,10pt]{IEEEtran}
\IEEEoverridecommandlockouts

\usepackage{cite}
\usepackage{amsmath,amssymb,amsfonts}
\usepackage{algorithmic}
\usepackage{graphicx}
\usepackage{textcomp}
\usepackage{xcolor}
\usepackage{xparse}

\usepackage{subcaption}
\usepackage{bm}
\usepackage{pgfplots}
\usepackage{xparse}
\usepackage[normalem]{ulem}
\usepackage{nicefrac}
\usepackage{csvsimple}
\usepackage{filecontents}
\usepackage{booktabs}
\usepackage{array}
\usepackage{ragged2e}
\usepackage{microtype}

\usepackage{etoolbox}
\newtoggle{usepdfs}

\togglefalse{usepdfs} 

\usepackage{tikz}
\usetikzlibrary{external}
\newcolumntype{P}[1]{>{\RaggedRight\arraybackslash}p{#1}} 
\newcolumntype{Q}[1]{>{\centering\arraybackslash}p{#1}}   

\usepackage[normalem]{ulem}  

\def\BibTeX{{\rm B\kern-.05em{\sc i\kern-.025em b}\kern-.08em
    T\kern-.1667em\lower.7ex\hbox{E}\kern-.125emX}}

\newcommand{\revised}[1]{#1}

\newcommand{\removewhitespace}{\vspace{-4mm}}

\makeatletter

\DeclareMathOperator{\Down}{D}
\DeclareMathOperator{\Up}{U}

\makeatother
\begin{document}

\title{EqDeepRx: Learning a Scalable and Interference Mitigating MIMO Receiver}

\author{
\IEEEauthorblockN{Mikko Honkala, Dani Korpi, Elias Raninen, and Janne M. J. Huttunen}\\
\IEEEauthorblockA{\textit{Nokia Bell Labs}\\
\textit{Espoo, Finland}\vspace{-10mm}}
}

\newcommand{\hop}{\mathsf{H}}
\newcommand{\bo}{\boldsymbol}
\newcommand{\complex}{\mathbb C}
\newcommand{\comp}{\mathbin{\circ}}
\newcommand{\Fro}{\mathrm{F}}
\newcommand{\diag}{\mathrm{diag}}

\maketitle

\begin{abstract}
While machine learning (ML)-based receiver algorithms have received a great deal of attention in the recent literature, they often suffer from poor scaling with increasing spatial multiplexing order and lack of explainability and generalization. This paper presents \emph{EqDeepRx}, a practical deep-learning-aided multiple-input multiple-output (MIMO) receiver, which is built by augmenting linear receiver processing with carefully engineered ML blocks. At the core of the receiver model is a shared-weight \emph{DetectorNN} that operates independently on each spatial stream or layer, enabling near-linear complexity scaling with respect to multiplexing order. To ensure better explainability and generalization, EqDeepRx retains conventional channel estimation and augments it with a lightweight \emph{DenoiseNN} that learns frequency-domain smoothing. To reduce the dimensionality of the DetectorNN inputs, the receiver utilizes two linear equalizers in parallel: a linear minimum mean-square error (LMMSE) equalizer with interference-plus-noise covariance estimation and a regularized zero-forcing (RZF) equalizer. The parallel equalized streams are jointly consumed by the DetectorNN,  after which a compact \emph{DemapperNN} produces bit log-likelihood ratios for channel decoding. 5G/6G-compliant end-to-end simulations across multiple channel scenarios, pilot patterns, and inter-cell interference conditions show improved error rate and spectral efficiency over a  conventional baseline, while maintaining low-complexity inference and support for different MIMO configurations without retraining.
\end{abstract}

\section{Introduction}

Machine learning (ML)-based enhancements to wireless communication systems have received a great deal of attention in the recent years among the research community. Primary motivation for such research has been to obtain computationally feasible algorithms for very complex tasks, or obtain new types of data-driven algorithms, by emulating them with deep neural networks. Specifically in the physical layer, it has been shown that ML can bring significant gains in terms of spectral efficiency \cite{oshea17, He19a, huang2020}. However, practical challenges such as memory and computational demands still remain and need to be adequately addressed before these techniques can be fully adopted in real networks.

In this paper, our focus is on ML-driven physical layer receiver processing. Previously, in \cite{Honkala21}, we proposed a deep learning orthogonal frequency-division multiplexing (OFDM) receiver DeepRx which was based on a convolutional neural network (CNN) architecture. In DeepRx, the CNN processes the received data for the whole transmission time interval (TTI) at once and outputs the log-likelihood ratios (LLRs) of the transmitted bits. The model was trained with single-input multiple-output (SIMO) data and it reached high performance compared to a traditional receiver in simulation-based evaluations. In \cite{korpi2021}, we showed the benefits of incorporating \emph{expert knowledge} into the architecture via layers of operations building on traditional receiver signal processing. This allowed the model to achieve competitive performance also with multiple-input and multiple-output (MIMO) scenarios, with reasonable computational complexity.

Herein, we extend our previous deep learning based receiver DeepRx \cite{Honkala21,korpi2021} toward a structure that is tailored for meeting the high MIMO multiplexing order and throughput demands of practical receiver deployment in real-world networks in 6G and beyond. This is done by expanding on the idea of using expert knowledge together with deep learning. The improved receiver, which we call \emph{EqDeepRx}, uses a novel architecture that combines parts of the conventional MIMO OFDM signal processing flow with neural network (NN) stages. A distinguishing feature of EqDeepRx is that it adopts a parallel equalizer (EQ) design that allows for reducing the computational complexity. As in \cite{korpi2021}, including EQs incorporates predefined operations for equalization into the NN model (e.g. matrix inversions) that would be difficult to approximate with traditional NN structures without excessively high complexity. In this work, we utilize this EQ structure even further and design a neural network architecture that is invariant to the number of spatially multiplexed MIMO transmissions (i.e., MIMO layers). This is a desirable feature in practical deployments to allow for flexible MIMO scheduling. Furthermore, we enable effective operation under strong inter-cell interference by utilizing an EQ specifically tailored for interference. Finally, we show that employing multiple parallel EQs can bring benefits both in performance and training stability.

The main contributions of this article are listed below.
\begin{itemize}
\item We propose a novel ML-based receiver, EqDeepRx, which combines conventional receiver processing with NN layers specifically designed to keep computational complexity reasonable. The defining characteristics of the EqDeepRx receiver model are: (i) introducing deep learning in the channel denoising and soft symbol detection phases, while relying on otherwise conventional receiver processing, (ii) taking advantage of two parallel equalizers, and (iii) scaling nearly linearly to support arbitrary number of MIMO layers, which also makes the receiver feasible for realistic systems.
\item As one of the EQs, we employ a \emph{linear minimum mean square error} (LMMSE) equalizer with \emph{interference-plus-noise covariance matrix} (INCM) estimation, allowing effective operation under strong inter-cell interference.
\item \revised{To show that proposed EqDeepRx model achieves state-of-the-art performance, the model is evaluated across a wide range of channel conditions and system scenarios, both in terms of raw uncoded bit error rate (BER), as well as the block error rate (BLER) after channel decoding.} The latter is also used to quantify the gain of the receiver in terms of spectral efficiency improvement. It is also demonstrated that the developed architecture can achieve good accuracy across different MIMO layer allocations without retraining.
\item A comprehensive ablation study is performed, where the developed primary EqDeepRx architecture is compared to various alternatives in terms of ML processing and/or computational complexity. \revised{The absolute computational complexity of the NN layers of the different EqDeepRx variants is also quantified and reported in terms of floating point operations (FLOPs).}
\end{itemize}

The rest of this article is organized as follows. In Section~\ref{sec:sys_model}, we describe the basic system model, and also define the baseline LMMSE receiver. Then, in Section~\ref{sec:eqdeeprx}, the proposed EqDeepRx model architecture is described in detail. After this, the simulation results, including the ablation study, are presented in Section~\ref{sec:sim_results}, while Section~\ref{sec:conc} concludes the article.

\subsection{Related work}

ML-based approaches have been successfully applied to many parts of the radio physical layer. Some of the more widely studied examples include channel estimation \cite{neumann18,he18}, signal compression and detection \cite{qin2019, zhao2018}, and coding \cite{kim2020}. There is also a wide body of literature focusing on designing and training a complete receiver based on, at least partially, deep learning. The works in \cite{ye18,gao18} introduced two different approaches for learning a single-input single-output (SISO) receiver from data, using either a fully learned approach \cite{ye18} or one where ML was combined with expert knowledge \cite{gao18}. The aforementioned DeepRx ML receiver was proposed in \cite{Honkala21}, consisting of a fully convolutional ResNet model. All of these approaches were reported to provide performance gains over the conventional baselines in SISO transmission.

After these initial findings, the work has focused on scaling ML-based receiver approaches to cover a wider range of practical scenarios, while addressing the computational complexity. The problem of learning a MIMO detector was analyzed in \cite{Samuel19a}, assuming perfect channel knowledge. In \cite{korpi2021}, DeepRx was extended for MIMO transmissions, where it was shown that in order to reach sufficient performance, the ML architecture should incorporate expert knowledge via additional processing blocks that resemble conventional receiver processing. Another ML-based receiver variant intended for MIMO OFDM systems was proposed in \cite{Cammerer23a}. It builds on a combination of convolutional layers and graph neural networks (GNNs) to build an ML-based OFDM receiver that can operate with different numbers of MIMO streams. \revised{This model is relying purely on neural network layers, with no elements from conventional receivers included between the OFDM demodulator and channel decoder. Performance gain over the baseline approach is reported with all considered MIMO configurations.}
Use of GNNs within a receiver algorithm has also been investigated in \cite{Clausius25a,Norouzi25a,Liu25a}. The work in \cite{Clausius25a} implements a joint detector and decoder with a GNN and demonstrates substantial performance gain over comparable baselines. In \cite{Norouzi25a}, GNN-based channel estimation for 5G is investigated, which is shown to slightly improve performance, while having significantly lower computational complexity. Another GNN-aided receiver approach is proposed in \cite{Liu25a}, where GNNs are utilized in the posterior estimation phase to compensate for approximation errors.

\revised{The key idea in our proposed EqDeepRx model is to combine traditional receiver components with machine learning models to reduce computational cost compared to fully machine-learning-based methods. A similar ML-based MIMO receiver was proposed in \cite{Goutay21a}, which, however, takes advantage of machine learning models in somewhat different manner. In their work CNNs are used for improving the channel estimation and symbol demapping. Particular attention is paid to improving the accuracy of the channel estimation error estimates, while channel estimate denoising is done with conventional methods. Performance improvement over the baseline approach is reported especially at high UE speeds. However, their work does not consider inter-cell interference nor is it included in their spatial channel-estimation error covariance model. In addition, their work partially addresses the scaling issues for large antenna arrays, although some parts of the channel estimator still require processing that does not scale linearly to antenna counts. This leaves some potential for further optimization.}

Most of the aforementioned works assume a linear system, where the ML-based algorithms are designed to outperform conventional receivers due to learning more accurate processing steps based on the training data distribution. However, significant research effort has also been focused on dealing with different hardware impairments by using deep learning. For instance, a DeepRx-type CNN receiver has been shown to be robust against power amplifier-induced nonlinear distortion when equipped with additional Fourier transforms \cite{pihlajasalo2023}. The work in \cite{Farhadi23a} demonstrated that ML-based demapper can be effective against such PA-induced distortion with discrete Fourier transform-spread OFDM (DFT-s-OFDM) waveforms.

\begin{figure*}[ht]
	\centering
	\includegraphics[width=\textwidth,trim={0 0 0 0},clip]{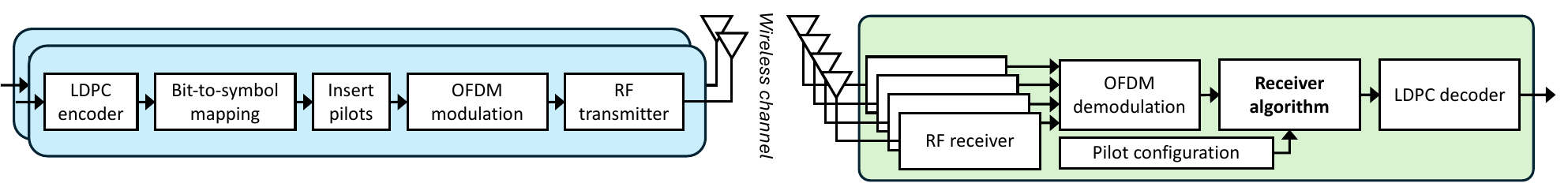}
	\caption{\label{fig:block_diag}Overall block diagram of the considered system. In this article, we consider different alternatives for the receiver algorithm part, including the conventional baseline as well as the proposed EqDeepRx ML receiver.}
	\removewhitespace
\end{figure*}

An interesting longer-term line of research has focused on deep learning based end-to-end solutions that simultaneously optimize both the transmitter and the receiver. There are multiple proposals where the transmitter-receiver link is learned from data without necessarily any pre-specified modulation scheme~\cite{cammerer2019,korpi2023,Lo25a,Cheng25a}. The early work focused on a fully learned approach with learned waveforms \cite{cammerer2019}, while the work has more recently focused on OFDM waveforms with some learned properties, such as constellation shape \cite{korpi2023,Lo25a,Cheng25a}. The main benefit of utilizing a learned constellation shape with a jointly learned ML receiver is that the system learns to communicate without any pilot signals, improving the spectral efficiency. More recently, the use of superimposed pilots has also been proposed as another approach for achieving similar spectral efficiency gain without necessarily having to modify the constellation shape \cite{Zou25a}. However, despite the higher performance compared to utilizing only an ML-based receiver, implementing such end-to-end approaches in a practical radio network requires significant modifications to existing wireless standards, which can delay their implementation in practical systems.

This paper builds on and extends DeepRx, making it a realistic ML-based receiver that meets the practical demands of future cellular networks. This is accomplished by augmenting the architecture with carefully selected expert knowledge, in contrast to some of the other ML receivers in the literature (e.g., \cite{Cammerer23a}). \revised{The test scenarios are designed to account for most factors that impact practical cellular radios, such as inter-cell interference, diverse propagation conditions, hardware impairments, and multiple configurations.} Moreover, we focus on evaluating the developed receiver under 5G/6G compliant OFDM waveforms, to ensure that the reported results reflect what is achievable in the 6G timeframe.

\section{System Model}
\label{sec:sys_model}

In this section we formulate the system model and outline the receiver processing steps that are used as a basis for the proposed EqDeepRx architecture. Figure~\ref{fig:block_diag} depicts the system model on a high level. Throughout the paper, we restrict our analysis to the uplink (UL) case, where one or several UEs transmit data to a base station (BS). However, most of the algorithms are applicable also to the downlink (DL) case.

Consider OFDM transmissions, where one or several UEs transmit a total of $N_T$ spatial streams, also referred to as MIMO layers, to a base station with $N_R$ RX antennas over the duration of one slot. We assume that the number of layers is equal to the number of transmit (TX) antennas. We denote the number of data carrying subcarriers by $N_F$ and the number of OFDM symbols within a slot by $N_S$ (typically $N_S = 14$). The baseband-equivalent model for the received signal $\mathbf{y}_{ij} \in \mathbb{C}^{N_R \times 1}$ corresponding to the $i$th subcarrier in the $j$th OFDM symbol can be expressed as
\begin{align}
	\mathbf{y}_{ij} = \mathbf{H}_{ij} \mathbf{x}_{ij} + \mathbf{v}_{ij} +\mathbf{n}_{ij}
    , %
	\label{eq:rx_signal}
\end{align}
where $\mathbf{H}_{ij} \in \mathbb{C}^{N_R \times N_T}$ is a deterministic (non-random) channel matrix, $\textbf{x}_{ij} \in \mathbb{C}^{N_T \times 1}$ comprise the transmitted data signals corresponding to the MIMO layers, $\mathbf{v}_{ij} \in \mathbb{C}^{N_R \times 1}$ constitutes the interfering signals, and $\mathbf{n}_{ij} \in \mathbb{C}^{N_R \times 1}$ is the noise. We often use the notation $\mathbf{d}_{ij} = \mathbf{v}_{ij} +\mathbf{n}_{ij}$ for the \emph{interference-plus-noise} vectors. We assume that $\mathbf{x}_{ij}$, $\mathbf{v}_{ij}$, and $\mathbf{n}_{ij}$ are mutually independent and zero mean random vectors.

{\bf Remark:} Note that the above frequency-domain model cannot express all physical time-domain channel phenomena such as inter-carrier interference (ICI) or inter-symbol interference (ISI). Hence, the interference term $\mathbf{v}_{ij}$ is assumed to only include inter-cell interference consisting of overlapping transmissions of UEs in adjacent cells. However, we would like to highlight that while the signal model used for the receiver design does not accommodate ISI and ICI, the training and validation data generation is carried out using a complete time-domain simulator. Therefore, both ISI and ICI effects are present in the simulations, and consequently the trained models can learn to mitigate these effects during training to the extent allowed by their model architectures and training procedures.

\subsection{Equalization}
\label{sec:lmmse}
\label{sec:equalization}

In EqDeepRx, we use two equalizers in parallel: the \emph{regularized zero forcing} (RZF) equalizer and the LMMSE equalizer. For the signal model \eqref{eq:rx_signal} (omitting the subscripts $ij$), the RZF equalizer is defined as
\begin{equation}
    \label{eq:RZF}
    \mathbf{W}_{\text{RZF}} = 
    (\mathbf{H}^\hop\mathbf{H} + \alpha\mathbf{I})^{-1}\mathbf{H}^\hop
    ,
\end{equation}
where $\alpha \mathbf{I}$ is a regularizing term. If $\alpha=0$, we obtain the traditional zero-forcing (ZF) equalizer. In the simulations, we use the empirically determined value $\alpha=10^{-4}$. 

In order to perform the equalization in practice, we need estimates of the channel matrix for each resource element $(i,j)\in \mathcal D$, where $\mathcal D$ denotes the set of data symbols. We provide detailed discussions regarding the channel estimation in Sec. \ref{sec:channel-estimation}. Given the channel estimates $\widehat{\mathbf H}_{ij}$, we plug them in to \eqref{eq:RZF} to obtain $\widehat{\mathbf W}_{ij,\text{RZF}}$. Since we desire unit-gain symbols, we further scale the equalizer by the inverse of $\mathbf D_{ij,\text{RZF}}=\diag(\widehat{\mathbf{W}}_{ij,\text{RZF}} \widehat{\mathbf H}_{ij})$, yielding the equalized symbols
\begin{equation}
    \widehat{\mathbf{x}}_{ij,\text{RZF}} = 
    \mathbf{D}_{ij,\text{RZF}}^{-1}
    \widehat{\mathbf{W}}_{ij,\text{RZF}}
    \mathbf{y}_{ij}
    , ~ (i,j) \in \mathcal{D}
    .
\end{equation}

As for the LMMSE equalizer, $\mathbf{W}_{\text{LMMSE}}$, it is defined as the linear filter that minimizes the mean squared error (MSE) between the estimated and transmitted symbols. For the signal model \eqref{eq:rx_signal} (omitting the subscripts $ij$), it is given by
\begin{equation}
\revised{\mathbf{W}_{\text{LMMSE}}
    = 
    \arg \min _{
         \mathbf{W} \in \complex^{N_T \times N_R}
    }
    \mathbb E[\Vert \mathbf{x}-\mathbf{W} \mathbf{y}\Vert^2]
    = \mathbf{R}_{\mathbf{xy}} \mathbf{R}_{\mathbf{y}}^{-1}},
\end{equation}
where
$
    \mathbf R_{\mathbf y}
    =
    \mathbb{E}[\mathbf y \mathbf y^\hop]
    = 
    \mathbf{H}\mathbf{R}_\mathbf{x}\mathbf{H}^\hop + \mathbf{R}
$,
$
    \mathbf R_{\mathbf x \mathbf y}
    =
    \mathbb{E}[\mathbf x \mathbf y^\hop]
    = 
    \mathbf{R}_\mathbf{x}\mathbf{H}^\hop
$, $\mathbf R_{\mathbf x} = \mathbb E[\mathbf x \mathbf x^\hop]$, and
\begin{equation}
    \mathbf{R} = \mathbb E[\mathbf d \mathbf d^\hop] = \mathbb E[(\mathbf v+ \mathbf n) (\mathbf{v}+\mathbf{n})^\hop ]
    \label{eq:INCM}
\end{equation}
is the INCM. As the constellation must be known both by the transmitter and the receiver, we may assume $\mathbf{R}_{\mathbf{x}} = \mathbf{I}$, and the equalizer simplifies to
\begin{align}
    \label{eq:LMMSE}
    \mathbf{W}_{\text{LMMSE}}
    &=
    \mathbf{H}^\hop (\mathbf{H}\mathbf{H}^\hop + \mathbf{R})^{-1}.
\end{align}

In a practical implementation, in addition to the the estimated channel matrix, we also need to estimate the INCM. However, instead of obtaining an estimate for each resource element, the INCM $\mathbf{R}$ is estimated over each \emph{interference coherence bandwidth} defined as the interval of subcarriers for which the interference statistics may be assumed to remain approximately constant. We provide a detailed discussion regarding the estimation of the INCM in Sec. \ref{sec:INCM-estimation}.

Assuming we have channel estimates $\widehat{\mathbf H}_{ij}$ and an estimate $\widehat{\mathbf R}$ of the INCM, we may plug in the estimates in \eqref{eq:LMMSE} to obtain $\widehat{\mathbf{W}}_{ij,\text{LMMSE}}$. In order to obtain unit-gain symbols, as before we further scale the equalizer by the inverse of $\mathbf{D}_{ij} = \diag(\widehat{\mathbf{W}}_{ij,\text{LMMSE}} \widehat{\mathbf{H}}_{ij})$, yielding the estimated symbols
\begin{align}
    \widehat{\mathbf{x}}_{ij,\text{LMMSE}} = 
    \mathbf{D}_{ij,\text{LMMSE}}^{-1}
    \widehat{\mathbf{W}}_{ij,\text{LMMSE}}
    \mathbf{y}_{ij}
    , ~ (i,j) \in \mathcal{B},
\end{align}
where $\mathcal{B}$ denotes the set of indices corresponding to REs carrying data symbols in an interference coherence bandwidth. 

{\bf Remark}: the LMMSE equalizer can also be expressed in the form $\mathbf{W}_{\text{LMMSE}}= (\mathbf{H}^\hop \mathbf{R}^{-1} \mathbf{H} + \mathbf I)^{-1} \mathbf{H}^\hop \mathbf{R}^{-1}$. Since $N_T\ll N_R$ and the inverse of $\mathbf{R}$ need to be calculated only once for each interference coherence bandwidth, this form is usually less computationally expensive. Also note, that given the alternative formulation of the LMMSE and the fact that we assumed $\mathbf{R}_{\mathbf x}=\mathbf I$, the only remaining difference between the RZF and the LMMSE is that $\mathbf R$ is replaced with $\alpha \mathbf I$.

\subsection{Channel estimation}
\label{sec:channel-estimation}

To perform equalization, we must first estimate the channel matrices $\mathbf{H}_{ij}$ for all subcarriers $i=1,\ldots,N_F$ and all OFDM symbols $j=1,\ldots,N_S$. To this end, pilot information or demodulation reference signals (DMRSs) that are \emph{a priori} known at the receiver constitute a portion of the transmitted data. The pilot information need to be multiplexed and processed so that decent estimates can be obtained for each layer (i.e., each column of the channel matrix). Regardless of the specific type of multiplexing, the receiver first calculates rank one raw channel estimates using
\begin{align}
	\widehat{\mathbf{H}}_{ij,\text{raw}} = \frac{1}{||\mathbf{x}_{ij}||_2^2} \mathbf{y}_{ij} \mathbf{x}_{ij}^\hop, \quad (i,j) \in \mathcal{P} ,  \label{eq:raw_channel_estimate}
\end{align}
where $\mathcal P = \{(i,j) : \mathbf{x}_{ij}~\text{contains pilot information}\}$ and $||\cdot||_2$ denotes the Euclidean norm. We assume pilot sequences that are orthogonal in time and frequency (i.e., no code domain multiplexing). Hence, a single pilot symbol in a specific RE is used to estimate a specific layer (column of the channel matrix) at that specific $(i,j)\in \mathcal P$, meaning that each pilot symbol $\mathbf{x}_{ij}$ only has one non-zero element and consequently the individual raw channel matrix estimates contain only one non-zero row. In order to obtain the final channel matrix estimates denoted by $\widehat{\mathbf{H}}_{ij}$ for all resource elements, we utilize linear interpolation followed by smoothing, or denoising, \revised{which is commonly carried out with} a static frequency-domain filter.

\subsection{Interference-plus-noise covariance matrix (INCM) estimation}
\label{sec:INCM-estimation}

\revised{The INCM estimate $\widehat{\mathbf{R}}$ is computed for each interference coherence bandwidth, which is an interval of subcarriers for which we may assume that the interference statistics remain approximately static. We use an interference coherence bandwidth corresponding to two PRBs, equaling 24 subcarriers.} 

Let $\mathcal B$ denote an index set of resource elements corresponding to a particular interference coherence bandwidth. \revised{The INCM estimate corresponding to $\mathcal B$ is computed from the interference-plus-noise vectors estimated from the pilot symbols in $\mathcal B$.} Hence, the interference-plus-noise vectors are estimated via
\begin{displaymath}
    \widehat{\mathbf{d}}_{ij} = \mathbf{y}_{ij} - \widehat{\mathbf{H}}_{ij} \mathbf{x}_{ij}, \quad (i,j) \in \mathcal{P} \cap \mathcal{B}.
\end{displaymath}
An initial estimate of the INCM $\mathbf{R}$ is obtained by computing the sample covariance matrix 
\begin{displaymath}
    \widehat{\mathbf{S}} =  \frac{1}{P} \sum_{i, j \in \mathcal{P} \cap \mathcal{B}} \widehat{\mathbf{d}}_{ij} \widehat{\mathbf{d}}_{ij}^\hop,
\end{displaymath}
where $P = |\mathcal P \cap \mathcal B|$, i.e., the number of pilot symbols in $\mathcal B$. 
The accuracy of the estimator is further improved using linear shrinkage estimation defined by 
\begin{equation}
    \widehat{\mathbf{S}}(\rho) = (1-\rho)\widehat{\mathbf{S}} + \rho
    [\mathrm{tr}(\widehat{\mathbf{S}})/N_{R}]\mathbf{I},~\rho\in [0,1]
    .
    \label{eq:OAS}
\end{equation}
\revised{
The shrinkage parameter $\rho$ is selected using the complex-valued version~\cite{raninenOracleApproximatingShrinkage2024} of the oracle approximating shrinkage (OAS) method~\cite{chenShrinkageAlgorithmsMMSE2010}, which approximates the MSE minimizing shrinkage parameter under the assumption of complex Gaussian disturbance. Note that although inter-cell interference is not strictly Gaussian because it originates from UE data transmissions in neighboring cells, the Gaussian assumption often provides a sufficiently accurate approximation.
}
\subsection{Demapping and Decoding}
\label{sec:demapping}

The final step in the physical layer receiver chain is to perform demapping, in which posterior probabilities of sent bits (soft bits) are calculated based on the symbol estimates $\widehat{\mathbf{x}}_{ij}$. The demapping is typically carried out by calculating the log-likelihood ratios (LLRs),
\begin{align}
	L_{ijkl} = \log \left( \frac{\operatorname{Pr}\left( c_l = 0 | \hat{x}_{ijk}\right)}{\operatorname{Pr}\left( c_l = 1 | \hat{x}_{ijk}\right)} \right), \quad l = 0,\hdots,B-1\label{eq:llr_exact}
\end{align}
where $\operatorname{Pr}\left( c_l = 0/1 | \hat{x}_{ijk}\right)$ is the conditional probability that the transmitted $l$th bit on $i$th subcarrier, $j$th OFDM symbol, and $k$th MIMO layer is $0/1$ given the observed symbol $\hat{x}_{ijk}$, and $B$ is the number of bits per symbol. For further details about the demapper, please refer to \cite{Honkala21}.

After demapping, the obtained LLRs are fed to the low-density parity-check (LDPC) decoder to compute final information bits from the estimated LLRs. In this work, we utilize 5G-compliant LDPC channel coding \cite{Hui18a}, but we omit a more detailed description for brevity.

\section{EqDeepRx}
\label{sec:eqdeeprx}

The original DeepRx architecture \cite{Honkala21} demonstrated significant gains in SIMO scenarios by jointly producing all LLRs from the received signal $\mathbf{y}$ and raw channel estimate $\widehat{\mathbf{H}}$, processing an entire slot with a ResNet-based CNN.  We attributed these gains to the network’s learned ability to exploit known symbol constellations and track both time- and frequency-selective fading.

When extending DeepRx to MIMO in \cite{korpi2021}, we first explored two variants:  
(i) a ResNet augmented with a maximum ratio combining (MRC) equalizer, and  
(ii) a fully-learned multiplicative transformation embedded within the network.  
Although both architectures improved upon practical baselines, they proved \revised{to be} suboptimal in practice—either incurring excessive filter counts and computational cost or failing to deal with inter-cell interference.  We attributed this to the fact that the neural network needed to learn and approximate operations similar to, for example, the matrix inversion to deal with inter-cell and inter-layer interference.

\begin{figure}[ht]
	\centering
	\includegraphics[width=\columnwidth]{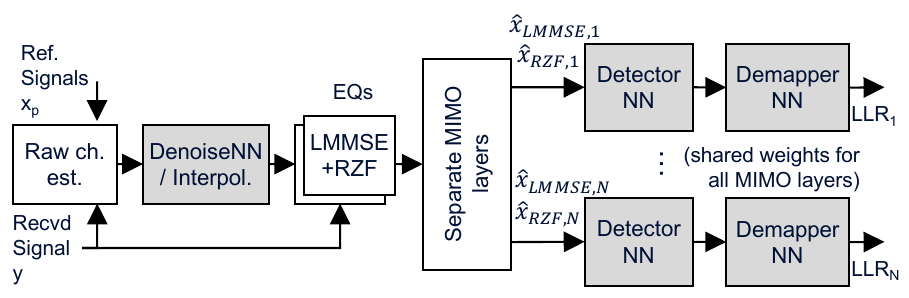}
    \caption{\label{fig:architecture}EqDeepRx high-level architecture. MIMO layers are handled in parallel separately after the EQs, allowing for efficient inference and layer count generalization. Blocks with gray background contain trainable parameters.}
\end{figure}

These findings motivate our hybrid \textbf{EqDeepRx} design shown in Fig.~\ref{fig:architecture}. EqDeepRx unites proven signal-processing components with a compact, per-MIMO-layer neural detector.  Our key design principles are:

\begin{itemize}
	\item \textbf{Hybrid channel estimation:} We retain DMRS-based raw channel estimation (Eq.~\eqref{eq:raw_channel_estimate}), followed by linear interpolation. \revised{Traditional frequency-domain smoothing is replaced by \emph{DenoiseNN}, a neural network trained to denoise raw channel estimates independently for each Tx-Rx antenna pair.} This mixture of deterministic and learned processing improves channel estimation quality while having a reasonable number of FLOPs.
	\item \textbf{Parallel equalization:} We apply both RZF and LMMSE equalizers in parallel to the denoised, interpolated channel estimate. \revised{Passing received signal through two equalizers enables the DetectorNN to leverage additional information about interference characteristics after the equalization stage: since the LMMSE and RZF equalizers differ only in the INCM term, their outputs are more similar when no interference is present and diverge substantially under strong interference. This dual-equalizer design alleviates the performance degradation caused by inaccuracies in the INCM estimate and yields more robust performance in high-interference scenarios. In essence, our dual-branch approach achieves superior interference suppression and noise-whitening while, by offloading equalization to classical algorithms, avoiding any inflation of the neural network's size.}
	\item \textbf{Per-MIMO-layer DetectorNN:} After equalization, MIMO layers are assumed to be separated.  A lightweight residual network—built from depthwise-separable 2D convolutions (alternating $1\times N$ and $N\times1$ kernels with subsampling)—then processes each layer independently.  By exploiting known constellation structure and tracking time–frequency variations, this DetectorNN delivers the bulk of the performance gains while scaling near-linearly in the number of layers.
\end{itemize}

As shown in Fig.~\ref{fig:architecture}, EqDeepRx takes as input the full‐TTI received signal \(\mathbf{y}\) and DMRS pilots, computes a channel estimate (Sec.~\ref{sec:channel-estimation}), applies parallel RZF/LMMSE equalizers (Sec.~\ref{sec:equalization}), then splits the per-layer outputs into the neural networks to produce final LLRs. Details are given in the following.

\subsection{Channel estimation and DenoiseNN}

In order to deal with varying channel characteristics and SNRs in the channel smoothing phase, we replace the conventional frequency-domain smoothing filter with a learned \emph{DenoiseNN}. As summarized in Table~\ref{tab:eqdeeprx_details}, DenoiseNN operates \emph{only at pilot positions} and is applied \emph{independently for each RX--TX antenna pair} $(r,t)$, i.e., for each complex pilot grid $\widehat{H}_{\text{raw}}^{(r,t)} \in \mathbb{C}^{F_P \times S_P}$ extracted from $\widehat{\mathbf{H}}_{\text{raw}}$. Note that $F_P$ and $S_P$ denote the number of pilot symbols along the subcarrier and OFDM symbol axes, respectively.

For each pair $(r,t)$, we first split complex values into two real channels (real/imag), yielding a tensor in $\mathbb{R}^{F_P \times S_P \times 2}$. This is processed by lightweight ResNet blocks applying down and upsampling for low-complexity smoothing / denoising operation (see \ref{sec:subsampled-blocks} for details). The blocks \revised{apply} 1D convolutions in the frequency direction while preserving per-symbol structure in time. 

To allow the denoiser to access all pilot symbols, we insert a lightweight \emph{symbol/time mixer} implemented as a shared \emph{pointwise ($1\times1$) convolution} after each ResNet block. Concretely, for each pilot subcarrier, we reshape the features so that the $S_P$ pilot symbols are stacked into the channel dimension (for a small subset $C_s$ of channels), apply a $1\times1$ conv to linearly mix these stacked features \emph{independently for each subcarrier} (i.e., with weights shared over frequency), and then reshape back. The remaining channels bypass the mixer, and a residual connection is used. See Table~\ref{tab:eqdeeprx_details} for detailed list of operations.

After denoising, we obtain the full-grid channel estimate $\widehat{\mathbf{H}} \in \mathbb{C}^{F \times S \times N_R \times N_T}$ by the same linear interpolation step as in the baseline, i.e., interpolating from the pilot grid $(F_P,S_P)$ to the full resource-element grid $(F,S)$. In summary, DenoiseNN learns the pilot-domain smoothing that would otherwise be implemented by a fixed low-pass filter, while keeping the remainder of the channel estimation pipeline conventional and lightweight. 

\subsection{Per-MIMO-Layer DetectorNN}
Equalization (Sec.~\ref{sec:lmmse}) with both LMMSE and RZF equalizers produces two complex tensors
\[
\widehat{\mathbf{X}}_{\mathrm{LMMSE}},\,\widehat{\mathbf{X}}_{\mathrm{RZF}}
\;\in\;
\mathbb{C}^{N_F\times N_S\times N_T}.
\]
For each MIMO layer $t$, we extract the $t$th slice of both tensors, concatenate along the channel axis, and split real/imaginary parts into separate channels to form
\[
\widetilde{\mathbf{X}}_t
\in
\mathbb{R}^{\,N_F\times N_S\times 4},
\]
which is the symbol input into the DetectorNN branch for layer~$t$.

In addition, we provide DetectorNN with explicit position information along the
frequency and OFDM-symbol axes by concatenating two fixed coordinate maps
(similar in spirit to \cite{Liu2018}). Specifically, we form
a normalized subcarrier-index map and a normalized symbol-index map,
\(\mathbf{M}_F,\mathbf{M}_S\in\mathbb{R}^{N_F\times N_S\times 1}\), with entries
\([\mathbf{M}_F]_{f,s}=2\frac{f}{N_F-1}-1\) and \([\mathbf{M}_S]_{f,s}=2\frac{s}{N_S-1}-1\) ($f=0,\ldots,N_F-1$, $s=0,\ldots,N_S-1$). Concatenating these maps to the
real/imaginary LMMSE/RZF channels yields the final DetectorNN input
\[
\widetilde{\mathbf{X}}^{\text{pos}}_t
=
\mathrm{concat}\!\left(\widetilde{\mathbf{X}}_t,\mathbf{M}_F,\mathbf{M}_S\right)
\in \mathbb{R}^{N_F\times N_S\times 6},
\]
allowing the convolutional filters to learn frequency-/time-position-dependent
processing when beneficial (for example DMRS symbol positions in time and input borders).

\begin{figure}[ht!]
	\centering
	\includegraphics[width=0.485\textwidth]{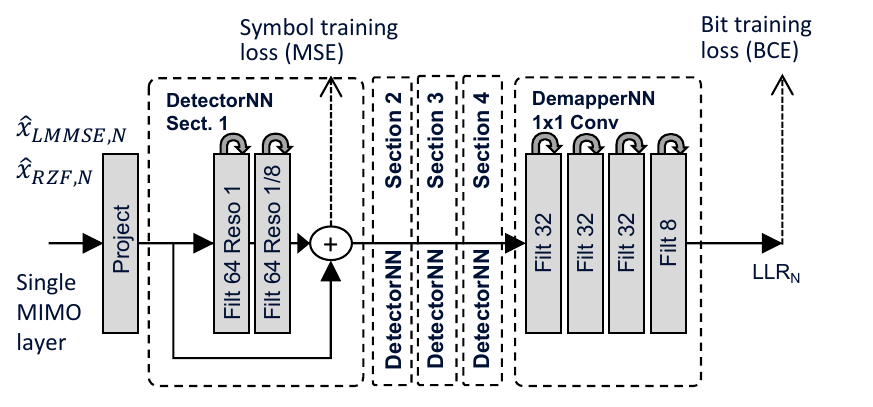}
	\caption{\label{fig:detectordemapper}Architecture of the DetectorNN and DemapperNN. This part of the model handles each MIMO layer separately in parallel.}
\end{figure}

DetectorNN, depicted in Fig. \ref{fig:detectordemapper}, processes each MIMO layer independently with shared weights.  For layer \(t\), the real‐valued input \(\widetilde{\mathbf{X}}^{\text{pos}}_t\in\mathbb{R}^{N_F\times N_S\times6}\) is first projected via a 1×1 2D convolution to produce
\[
\widetilde{\mathbf{X}}_t^{(0)}\in\mathbb{R}^{N_F\times N_S\times C},
\]
where \(C\) is the network’s channel count.  While this projection could be avoided via skip-connection projections, it makes the setup easier and does not affect link performance. The network then consists of \(K\) sections (typically \(K=4\)), each comprising two residual blocks described in Sec. \ref{sec:subsampled-blocks} at full ($N=1$) and one‐eighth resolution  ($N=8$).

Each section \(k\) also applies a section-level shortcut: if \(\widetilde{\mathbf{X}}_t^{(k-1)}\) is the input to the section and \(\mathbf{Z}_t^{(k)}\) is the updated tensor after applying the residual blocks, the section‐level output is then $\widetilde{\mathbf{X}}_t^{(k)}
=\widetilde{\mathbf{X}}_t^{(k-1)}+\mathbf{Z}_t^{(k)}$ (see Figure \ref{fig:detectordemapper}).

\subsection{Subsampled Residual Blocks}
\label{sec:subsampled-blocks}

Inspired by \cite{Peng2018}, we reduce the computational complexity of the residual networks by subsampling some of the residual blocks, while keeping the residual path in full resolution. In addition, we separate 2D convolutions into two consecutive 1D convolutions, similar to \cite{Jin2014}.

\begin{figure}[htb]
  \centering
  \includegraphics[width=0.49\textwidth, trim={0 16cm 0 0}, clip]{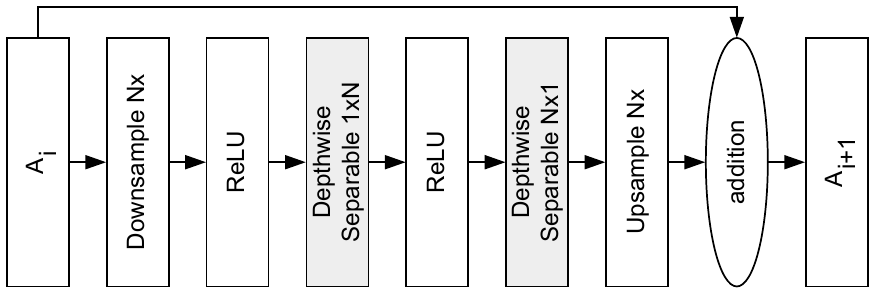}
  \caption{\label{fig:downsampling} The subsampling residual block for DetectorNN. DenoiseNN has similar structure, but both convolutions have Nx1 kernels for frequency-only operation.}
\end{figure}

Figure~\ref{fig:downsampling} depicts the subsampling residual block used in DetectorNN and DenoiseNN.  
Let $\Down$ denote the $N{:}1$ down-sampling operator and $\Up$ its inverse.  The block implements
\begin{equation}
	A_{i+1} = A_i + (\Up\comp f\comp \Down)(A_i),
	\label{eq:resblock}
\end{equation}
with
\begin{equation}
	f = 
	\mathrm{DW}_{13\times1}\comp
	\mathrm{ReLU}\comp
	\mathrm{DW}_{1\times13}\comp
	\mathrm{ReLU}
	\label{eq:fsubnet}
\end{equation}
using two depthwise-separable 2D convolutions with asymmetric $1\times13$ and $13\times1$ kernels.  This design preserves detailed information from each RE via the residual connection, while keeping the number of FLOPs low. We use a simple nearest-neighbour subsampling strategy both in down and upsampling. Note that when the input and output channel counts differ, we use the standard approach of projecting the skip (residual) branch with a $1\times1$ (pointwise) convolution.

\subsection{DemapperNN}

DemapperNN (see Figure \ref{fig:detectordemapper}) is a learned demapper that processes each MIMO layer and resource element independently. It begins with the real‐valued frequency–time feature map \(\widetilde{\mathbf{X}}^{(K)}\) produced by DetectorNN and then applies four standard ResNet‐style residual blocks—each using two 1×1 convolutions and ReLU activations—into bit‐wise LLRs. Note that 1x1 convolution corresponds to a fully connected layer applied to each RE independently (and via the global architecture, to each MIMO layer independently).

Finally, the output of the fourth and final residual block, denoted by \(D^{(4)}\), already has channel dimension \(B\), i.e., one channel per bit. Therefore, no additional output projection is required, and the per‐bit log‐likelihood ratios are read out directly as
\[
\mathrm{LLR}_b = \bigl[D^{(4)}\bigr]_b,
\quad b = 1,\dots,B.
\]
Note that, similar to \cite{Honkala21}, we use $B=8$ to support a maximum modulation order of 256-QAM and extract only the required number of LLRs when using lower modulation orders.

\subsection{Training EqDeepRx}

Inspired by \cite{Honkala21,korpi2023}, training of EqDeepRx is carried out by optimizing the weighted cross‐entropy loss extended with a symbol loss. We define the total loss for the $q$th sample as follows:

\begin{equation}
	\begin{aligned}
L_q(\theta)
= \log_2\bigl(1+\mathrm{snr}_q\bigr)\,
\Bigl(
\underbrace{\mathrm{CE}_q(\theta)\vphantom{\sum_{k=1}^K\|\widehat{\mathbf{X}}_q^{(k)}(\bm{\theta})-\mathbf{X}_q\|_Fro^2}}
_{L_{\mathrm{bit},q}}
+\,
\underbrace{\lambda\sum_{k=1}^K\|\widehat{\mathbf{X}}_q^{(k)}(\bm{\theta})-\mathbf{X}_q\|_2^2}
_{L_{\mathrm{sym},q}}
\Bigr).
\end{aligned}
\end{equation}
where $\bm{\theta}$ denotes the set of trainable parameters, $\mathrm{CE}_q(\cdot)$ denotes binary cross entropy, $\mathrm{snr}_q$ is the linear signal-to-noise ratio (SNR) of the sample, and $\lambda$ is a tunable parameter. \revised{In the training, we used an experimentally chosen $\lambda=10^{-5}$}.
For the symbol loss, $\mathbf{X}_q$ and \(\widehat{\mathbf{X}}_q^{(k)}\) are the ground truth and the predicted symbol after each state \(k\), respectively. The latter is chosen to be the first two channels of the DetectorNN output (after the residual sum) that are reinterpreted as the real and imaginary parts (the remaining channels carrying auxiliary features).

\revised{{\bf Remark:} In practice, the effect of the symbol loss can vary between evaluation scenarios. In some evaluation scenarios the value of $\lambda$ can have an insignificant effect on the final accuracy of the trained model.  Nevertheless, it is still incorporated as part of the overall loss function since it has been observed to improve the model convergence with some hyperparameter configurations and under certain channel conditions. Moreover, it helps in ensuring that the output of the DetectorNN actually represents symbol estimates, which makes the EqDeepRx model more explainable, which can be a desired feature in some applications.}

To improve training stability, several approaches, such as batch normalization layers, can be used. In this work, we chose to regularize the per-channel batch–spatial mean and variance toward fixed targets,
\[
\mathcal{L}_{\text{mVCL}}
=\alpha\,\frac{\|\hat\mu-\mu_*\|_2^2+\|\hat v-\sigma_*^2\|_2^2}{C},
\]
where $\hat\mu,\hat v$ are per-channel mean/variance (over batch and spatial dimensions), with  $\mu_*=0$, $\sigma_*^2=1$, and $\alpha=10^{-5}$ and $C$ the number of channels in a layer. This is a simplified variant of VCL~\cite{Littwin2018}. However, we have experimentally noticed that our results are not dependent on this specific choice and similar results can be reached with normalizations. 

\section{Simulation Results}
\label{sec:sim_results}

The training and performance evaluations of the proposed EqDeepRx ML receiver are carried out using Sionna open source library  \cite{sionna}. In particular, we have simulated the system depicted in Fig.~\ref{fig:block_diag} to first produce training data in an online fashion, after which we have validated the models under the chosen scenarios. Note that the online training data generation prevents overfitting by generating novel training data samples for each training batch. The detailed configuration of EqDeepRx is described in Table~\ref{tab:eqdeeprx_details}, which represents the primary model architecture and is used unless stated otherwise. For training, we used a batch size of 112, initial learning rate of $4.4 \times 10^{-3}$ for the LAMB optimizer, linear learning rate decay to zero, and $\sim$70k iterations (training consisted of $\sim$8M samples). Table~\ref{tab:sim_param} collects the key training and validation parameters, used in the forthcoming results unless stated otherwise.

\begin{table*}[htbp]
	\centering
	\scriptsize
	\setlength{\tabcolsep}{2pt} 
	\renewcommand{\arraystretch}{1.15}
	\caption{Details of DenoiseNN, DetectorNN and DemapNN with default configuration. DenoiseNN operates per antenna pair, while DetectorNN and DemapperNN operate per MIMO layer. Note that depth refers to the number of ResNet blocks.}

	\begin{tabular}{P{0.08\linewidth} P{0.17\linewidth} P{0.27\linewidth} P{0.12\linewidth} P{0.07\linewidth} Q{0.05\linewidth} P{0.06\linewidth} P{0.08\linewidth}}
		\toprule
		\textbf{Section} & \textbf{Layer} & \textbf{Description} & \textbf{Output shape} & \textbf{Filters} & \textbf{Depth} & \textbf{Subsample} & \textbf{Kernel sizes} \\
		\midrule
		\textbf{Channel est.} & \textbf{Input (H at pilots)} &
		Channel matrix at pilot pos. &
		\((F_P,S_P,N_R,N_T)\) & -- & -- & -- & -- \\
		& \textbf{ForEachAntennapair} &
		Each antenna pair separately over \(N_R,N_T\) &
		\((F_P,S_P,1)\) & -- & -- & -- & -- \\
		& \textbf{Complex\(\rightarrow\)Real} &
		Convert complex to two real channels &
		\((F_P,S_P,2)\) & -- & -- & -- & -- \\
		& \textbf{DenoiseNN} &
		ResNet stack w/ time mixing &
		\((F_P,S_P,2)\) & \([64,64,64,2]\) & \textbf{4} &
		1,4,2,1 &
		\(13{\times}1\) \\
		& \textbf{Real\(\rightarrow\)Complex} &
		Convert real back to complex &
		\((F_P,S_P,1)\) & -- & -- & -- & -- \\
		& \textbf{CombineAntennas} &
		Reshape per-pair outputs back to \((N_R,N_T)\) &
		\((F_P,S_P,N_R,N_T)\) & -- & -- & -- & -- \\
		& \textbf{Interpolate pilots to full} &
		Interpolate from \((F_P,S_P)\) to full grid \((F,S)\) &
		\((F,S,N_R,N_T)\) & -- & -- & -- & -- \\
		\midrule
		\textbf{Equalizers} & \textbf{LMMSE \(\mid\) RZF } &
		Inputs: denoised \(H\) \((F,S,N_R,N_T)\) and \(Y\) \((F,S,N_R)\); stack \(\hat{X}\) across \(E{=}2\) &
		\((F,S,N_T,E{=}2)\) & -- & -- & -- & -- \\
		\midrule
		\textbf{DetectorNN} & \textbf{ForEachLayer (over \(N_T\))} &
		Process each MIMO layer separately \(N_T\) &
		\((F,S,2)\) & -- & -- & -- & -- \\
		& \textbf{Complex\(\rightarrow\)Real} &
		Convert complex streams to real &
		\((F,S,4)\) & -- & -- & -- & -- \\
		& \textbf{ConcatPosMaps} &
		Concat frequency and symbol coordinate maps &
		\((F,S,6)\) & -- & -- & -- & -- \\
		& \textbf{Project} &
		Linear 1x1 conv projection &
		\((F,S,64)\) & -- & -- & -- & -- \\
		& \textbf{Section \textbf{1/4}} & ResNet stack and section shortcut&
		\((F,S,64)\) & \([64,64]\) & \textbf{2} &
		1,8 & \(13{\times}1,1{\times}13 \) \\
		& \textbf{Section \textbf{2/4}} & ResNet stack and section shortcut &
		\((F,S,64)\) & \([64,64]\) & \textbf{2} &
		1,8 & \(13{\times}1,1{\times}13\) \\
		& \textbf{Section \textbf{3/4}} & ResNet stack and section shortcut &
		\((F,S,64)\) & \([64,64]\) & \textbf{2} &
		1,8 & \(13{\times}1,1{\times}13\) \\
		& \textbf{Section \textbf{4/4}} & ResNet stack and section shortcut &
		\((F,S,64)\) & \([64,64]\) & \textbf{2} &
		1,8 & \(13{\times}1,1{\times}13\) \\

		& \textbf{Output (per-layer symbols)} &
		Refined per-layer symbol estimates &
		\((F,S,64)\) & -- & -- & -- & -- \\
		\midrule
		\textbf{DemapperNN} & \textbf{ResNet stack} & Output LLR estimates per bit for single layer &
		\((F,S,B)\) & \([32,32,32,B]\) & \textbf{4} &
		1,1,1,1 &
		\(1{\times}1\) \\
		\midrule
		\textbf{Output} & \textbf{CombineMIMOLayers} &
		Stack/gather per-layer LLRs back along \(N_T\) &
		\((F,S,N_T,B)\) & -- & -- & -- & -- \\

		\bottomrule
	\end{tabular}

	\vspace{4pt}
	\emph{Dimensions:} \(F\) = 192 (subcarriers), \(S\) = 14 (OFDM symbols), \(N_R\) = 16 (RX antennas), \(N_T\) = 4 (MIMO layers), \(B\) = 8 (bits), \(E=2\) Eqs. Batch dim. omitted for clarity.
    \label{tab:eqdeeprx_details}
		\removewhitespace
\end{table*}

\begin{table}[t]
\setlength{\tabcolsep}{2pt}
\centering
\caption{Key simulation parameters of the numerical experiments. $\mathcal{U}$ denotes uniform distribution.}
\label{tab:sim_param}
\begin{tabular}{|c|c|c|}
\hline
\textbf{Parameter} & \textbf{Training} & \textbf{Validation}\\
\hline
Channel model & UMa & \shortstack{\rule{0pt}{2.4ex}UMa, CDL-C,\\CDL-D, UMi} \\
\hline
Waveform & \multicolumn{2}{c|}{OFDM, 30 kHz SCS}\\
\hline
No. of subcarriers & \multicolumn{2}{c|}{192} \\
\hline
No. of OFDM symbols per slot & \multicolumn{2}{c|}{14} \\
\hline
No. of RX antennas & \multicolumn{2}{c|}{16} \\
\hline
No. of MIMO layers & \multicolumn{2}{c|}{2--4} \\
\hline
No. of UE TX antennas & \multicolumn{2}{c|}{1} \\
\hline
SNR & $\mathcal{U}\left(0 \text{ dB}, 45 \text{ dB}\right)$ & Varied \\
\hline
INR & \multicolumn{2}{c|}{$\mathrm{Lognormal}\left(10 \text{ dB}, 5 \text{ dB}\right)$}\\
\hline
No. of interfering UEs & 0 or 1 & 0 or 1 \\
\hline
UE speed & $\mathcal{U}\left(0 \text{ m/s}, 35 \text{ m/s}\right)$ & Varied\\
\hline
Delay spread & UMa &  10--1100 ns\\
\hline
Modulation order &  \multicolumn{2}{c|}{64-QAM / 16-QAM (one experiment) }\\
\hline
Code rate & N/A & MCS Table 2\\
\hline
DMRS configuration & \multicolumn{2}{c|}{1 or 2 OFDM pilot symbols per slot}\\
\hline
\end{tabular}
\removewhitespace
\end{table}

The EqDeepRx model was trained using the Urban Macro (UMa) channel model, as defined in 3GPP TR 38.901 \cite{NR_38901}, using a wide range of SNRs and UE speeds. This channel model was adopted for training due to its randomized nature, for example, in terms of channel frequency selectivity and angular characteristics. \revised{Performance validations additionally use the CDL-C, CDL-D, and UMi channel models, all assuming 3GPP 38.901 antenna patterns \cite{sionna,NR_38901}.} Moreover, the model is trained without any error correction code, but when validating, low-density parity check (LDPC) encoding and decoding is used, in accordance with 5G specifications. \revised{We focus on 64-QAM, except for one 16-QAM experiment; each model is trained for its own modulation order, since a joint modulation-based, bit-masked model would not be optimal in terms of complexity. We use MCS Table 2 \cite{NR_38214}, with modulation orders corresponding to MCS indices 9--19 (code rates 0.46--0.85).} The used DMRS configuration is fully orthogonal across a maximum of four UEs. This is done by allocating the pilot symbols of each MIMO layer on every fourth subcarrier within a physical resource block (PRB), with staggered offsets. A similar allocation is used on all DMRS-carrying OFDM symbols.

The interference is modeled by generating a transmit signal from one interfering UE with a random timing offset. The signal of the interfering UE experiences an independently randomized propagation channel, and the interfering power is determined by a log-normally distributed interference-to-noise ratio (INR). This ratio determines what is the relative power of the interfering signal with respect to the receiver noise power. Note that in the forthcoming figures, the SINR is calculated based on the combined effect of the random interference-plus-noise power realizations for each slot. The figures are generated by dividing the validation samples into a predefined number of bins based on the SINR realizations and averaging the ensuing quantities within these bins. For each validation, we used 32k samples. Lastly, it should be noted that, for most of the results of EqDeepRx throughout this paper, we use the same trained model. The same single model, trained only with UMa, handles both of the utilized demodulation reference signal (DMRS) configurations and is evaluated across all considered channel models (CDL, UMa, and UMi) without retraining. \revised{Exceptions are the no-interference, 16-QAM, FO, ablation/sensitivity, and UE-count studies, which use separately trained models.}

\subsection{Bit and Block Error Rate}

\begin{figure}[!t]
	\centering
     \includegraphics{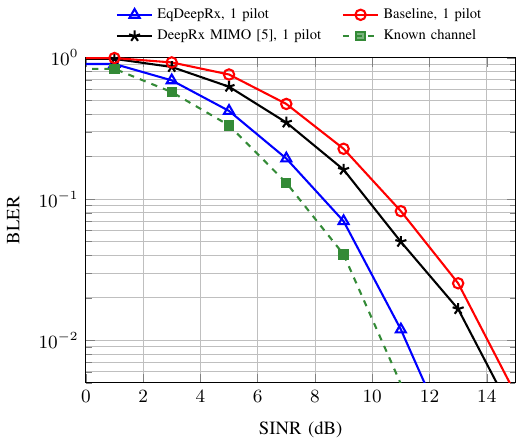} 
   	\caption{\revised{BLER with CDL-C and UE speed of 10--15\,m/s, no interference.}}
	\label{fig:mimo_study}
\end{figure}

\newcounter{plotblercount}
\NewDocumentCommand{\plotbler}{s m O{BLER UMa speed 10--15 m/s.} O{fig:bler} O{BLER} O{5e-3} O{23} O{-5} o}{
	\IfBooleanTF{#1}{\begin{subfigure}[t]{0.48\textwidth}}{\begin{figure}[!t]}
			\centering
                \stepcounter{plotblercount}
                \includegraphics{figs/bler\arabic{plotblercount}.pdf}
           		\caption{#3}
			\label{#4}
			\IfBooleanTF{#1}{\end{subfigure}}{\removewhitespace\end{figure}}%
}

\begin{figure*}[!t]
	\centering
	\begin{subfigure}[t]{0.48\textwidth}
			\centering
                \stepcounter{plotblercount}
                \includegraphics{figs/bler_rev\arabic{plotblercount}.pdf} \removewhitespace
			\caption{Uncoded BER, 10--15 m/s}
			\label{fig:ber_cdlc_15}
			\end{subfigure}
	\hfill
	\begin{subfigure}[t]{0.48\textwidth}
			\centering
                \stepcounter{plotblercount}
                \includegraphics{figs/bler_rev\arabic{plotblercount}.pdf}
								\removewhitespace
           
			\caption{BLER, 10--15 m/s}
			\label{fig:bler_cdlc_15}
			\end{subfigure}
	\caption{Simulated (a) uncoded BER and (b) BLER with CDL-C, under speed of 10--15 m/s. Data rate matched MCSs.}
	\label{fig:cdl_c_10_15ms}
	\removewhitespace
\end{figure*}

\begin{figure*}[!t]
	\centering
\plotbler*{plot_data/ver3/MUMIMO16x4_ch_uma_interf_64QAM_deeprx_standalone_denoise_12p_4UE_pre3x64_dil142_highDS_8gpu_lr4e_05_newint2/val_32000_CDLC_velo0_5_mcss11_12_int1_v8_1intfixcsv_data/mod06bler_bin_plot_per_pilot.csv}[BLER, 0--5 m/s.][fig:bler_cdlc_5][BLER][5e-3][13][-5][plot_data/ver3/MUMIMO16x4_interf_64QAM_TFIRC_1p_4UE/val_16000_CDLC_velo0_5_TFIRC_mcss11_12_int1_v8_1intfixcsv_data/mod06bler_bin_plot_per_pilot.csv]
	\hfill
	\plotbler*{plot_data/ver3/MUMIMO16x4_ch_uma_interf_64QAM_deeprx_standalone_denoise_12p_4UE_pre3x64_dil142_highDS_8gpu_lr4e_05_newint2/val_32000_CDLC_velo30_35_mcss11_12_int1_v8_1intfixcsv_data/mod06bler_bin_plot_per_pilot.csv}[BLER, 30--35 m/s.][fig:bler_cdlc_35][BLER][5e-3][15][-5][plot_data/ver3/MUMIMO16x4_interf_64QAM_TFIRC_1p_4UE/val_16000_CDLC_velo30_35_TFIRC_mcss11_12_int1_v8_1intfixcsv_data/mod06bler_bin_plot_per_pilot.csv]
	\caption{Simulated BLER with CDL-C, under speeds of (a) 0--5 m/s and (b) 30--35 m/s. Data rate matched MCSs.}
	\label{fig:cdl_c_0_5ms_and_30_35ms}
	\removewhitespace
\end{figure*}

Let us first analyze the EqDeepRx performance \revised{with 64-QAM modulation order} in terms of bit error rate (BER) and block error rate (BLER). The former is calculated directly from the model output, using the encoded transmit bit sequence as the ground truth values. The latter is determined based on the LDPC output, where a block is considered to have been correctly received if the decoded bit sequence matches with the transmitted information bits. The amount of information bits is aligned between the two DMRS configurations by using a larger MCS when there are 2 DMRS symbols. In the following results, we use MCSs 11 and 12 for slots with 1 DMRS and 2 DMRS, respectively. Since each slot contains 14 OFDM symbols in total, this means that the ensuing data rates are nearly identical (the slots with 1 DMRS carry 2.4\% less information bits). With this, the different DMRS configurations can be compared within the same BLER plot, since the lower MCS of the 1-DMRS transmissions results in higher redundancy and better decoding performance, at the cost of a less accurate channel estimate.

\revised{The baseline receiver is utilizing classical signal processing with LS channel estimation, static frequency-domain smoothing, and LMMSE equalization, as described in Section~\ref{sec:sys_model}}. In addition, \revised{the results include} a baseline with full channel knowledge. \revised{The BLER for this baseline corresponds to} the smaller of the two MCS indices.

\revised{For brevity, only CDL-C is shown across multiple UE speed ranges; CDL-D, UMa, and UMi use a single speed range.} This is because the relevant conclusions are the same across all the tested channel models.

\revised{We first consider a scenario where evaluations were performed without any inter-cell-interference. The corresponding results are shown in Fig.~\ref{fig:mimo_study}, where only the BLERs with 1 pilot are shown for clarity (all solutions achieved better BLER with 1 pilot in this scenario). In addition, the plot shows the performance achieved by the previous DeepRx MIMO model presented in \cite{korpi2021}, developed for interference-free scenarios. The implementation follows the original architecture, with the exception of replacing the maximum ratio combining (MRC) equalizer with regularized zero forcing (RZF) to make the DeepRx MIMO more competitive under the more spatially diverse CDL and UMa channel models. It can be observed that the EqDeepRx architecture performs well also in the interference-free case, owing to the use of the dual equalizers and shrinkage parameter. Moreover, it outperforms the prior-art solution from \cite{korpi2021} while having clearly lower computational complexity, as it relies more heavily on efficient predetermined receiver operations and has more optimized CNN architecture.}

\revised{In the remainder of this section, all the presented evaluations include inter-cell interference.} Next, Figs.~\ref{fig:cdl_c_10_15ms} and \ref{fig:cdl_c_0_5ms_and_30_35ms} show the performance of the different solutions with CDL-C channel model, across three different speed ranges. The uncoded BER before channel decoding is shown for the UE speed range of 10--15 m/s in Fig.~\ref{fig:cdl_c_10_15ms}, while only BLER is shown for the other speed ranges in Fig.~\ref{fig:cdl_c_0_5ms_and_30_35ms}, for brevity. Focusing first on the scenario where UE speed is 10--15 m/s, it can be observed that EqDeepRx outperforms the baseline both in terms of uncoded BER and BLER. The former can be considered a measure of the accuracy of the hard bit decisions at the model output, while the latter factors in also the effect of the LLR magnitudes as well as the reduced overhead of the 1 DMRS configuration. With the chosen rather low MCS levels, EqDeepRx achieves nearly similar BLER performance with the 1 pilot and 2 pilot configurations. \revised{Assuming a system-level BLER target of 10\%, the gain provided by EqDeepRx over the non-ML baseline is in the order of 2 dB in this scenario.}

\revised{The performance results in Fig.~\ref{fig:cdl_c_10_15ms} include also a receiver that is only utilizing the DenoiseNN for channel estimation, while the rest of the receiver processing relies on conventional methods. This can be considered to be representative of a state-of-the-art ML-based channel denoising algorithm, providing an useful benchmark for EqDeepRx. The DenoiseNN-only model was still trained end-to-end with the same loss function as the EqDeepRx and it used a traditional LMMSE equalizer and demapper.
 It can be observed that the DenoiseNN can improve on the baseline, while falling short of the performance of EqDeepRx. In this example scenario, DenoiseNN provides an improvement of approximately 1~dB, and EqDeepRx adds another 1~dB on top. For brevity, we include the DenoiseNN performance figures only for the 2 pilot option, with which it achieved higher performance.}

Similar performance gains are observed with the low and high speed scenarios in Fig.~\ref{fig:cdl_c_0_5ms_and_30_35ms}. With very low mobility, both the EqDeepRx and the baseline achieve the best performance with 1 DMRS, while in the high velocity case the baseline requires 2 DMRSs to achieve the typical operating point of 10\% BLER. EqDeepRx achieves identical BLER both with 1 and 2 DMRSs, indicating that it is not very susceptible to channel aging even with high UE speeds. The somewhat worse LLR quality is compensated by the higher redundancy of the lower MCS index. 
\revised{It should be noted that the apparent increase in BLER for the single-pilot baseline at higher SINRs may stem from estimation errors in the INCM that may be systematic in nature. Such estimation errors can be caused by, for example, imperfect noise cancellation in smoothing of the channel estimate. With lower SINRs, these errors usually do not have influence as those are hidden under noise and are covered by the additional noise covariance.}

\plotbler{plot_data/ver3/MUMIMO16x4_ch_uma_interf_64QAM_deeprx_standalone_denoise_12p_4UE_pre3x64_dil142_highDS_8gpu_lr4e_05_newint2/val_32000_CDLC_velo10_15_mcssfix11_12_int1_ds1e-08_1e-07_v8_1intfixcsv_data/mod06bler_bin_plot_per_pilot.csv}
[BLER with CDL-C, UE speed of 10--15 m/s, and RMS delay spread of 10--100 ns.]
[fig:bler_cdlc_10_15_ds1e-07]
[BLER][5e-3][13][-5]

To obtain further insight into how the EqDeepRx model performs under different channel conditions, Figs. \ref{fig:bler_cdlc_10_15_ds1e-07}, \ref{fig:bler_cdld_10_15_ds1e-07} and \ref{fig:bler_uma_15} show the BLER validation results for three additional channel scenarios: CDL-C and CDL-D with a maximum root mean square (RMS) delay spread of 100~ns, and UMa channel model. Each scenario is simulated with UE speeds of 10--15 m/s.

Investigating first the CDL-C simulation results with the reduced delay spread in Fig.~\ref{fig:bler_cdlc_10_15_ds1e-07}, we can observe good alignment with the higher delay spread scenario in Fig.~\ref{fig:cdl_c_10_15ms}. This indicates that the performance gain provided by EqDeepRx is not restricted to only very frequency selective channels.

\revised{This observation is further confirmed by the results obtained with the CDL-D channel model in Fig.~\ref{fig:bler_cdld_10_15_ds1e-07}, which represents a line-of-sight scenario with a maximum RMS delay spread of 100~ns.} This scenario seems to favor the 1 DMRS case to some extent more since both the EqDeepRx and the baseline achieve a higher performance with just 1 pilot. The gain provided by EqDeepRx is again in the order of 2~dB, similar to what was observed under the non-line-of-sight scenario in CDL-C.

\plotbler{plot_data/ver3/MUMIMO16x4_ch_uma_interf_64QAM_deeprx_standalone_denoise_12p_4UE_pre3x64_dil142_highDS_8gpu_lr4e_05_newint2/val_32000_CDLD_velo10_15_mcssfix11_12_int1_ds1e-08_1e-07_v8_1intfixcsv_data/mod06bler_bin_plot_per_pilot.csv}
[BLER with CDL-D, UE speed of 10--15 m/s, and RMS delay spread of 10--100 ns.][fig:bler_cdld_10_15_ds1e-07][BLER][5e-3][23][2]

\revised{Next, the achieved BLER under the UMa channel model is shown in Fig.~\ref{fig:bler_uma_15}, where the overall gain provided by EqDeepRx is to some extent higher than in the CDL scenarios.} Looking again at the BLER value of 10\%, EqDeepRx yields a gain of more than 4~dB. One explanation for this is the fact that EqDeepRx was also trained with the UMa channel data. Despite the random nature of this channel model, there are still opportunities for data-driven receivers to optimize their behaviour for the prevailing data distribution. Moreover, similar to CDL-C, there is practically no difference in the performance of the receivers with 1 DMRS and 2 DMRSs, indicating that in this scenario the additional redundancy provided by the reduced overhead is comparable to a more accurate channel estimate.

\plotbler{plot_data/ver3/MUMIMO16x4_ch_uma_interf_64QAM_deeprx_standalone_denoise_12p_4UE_pre3x64_dil142_highDS_8gpu_lr4e_05_newint2/val_32000_CDLuma_velo10_15_mcss11_12_int1_newint_16bins_v5csv_data/mod06bler_bin_plot_per_pilot.csv}[BLER with UMa and UE speed of 10--15 m/s. Data rate matched MCSs.][fig:bler_uma_15][BLER][5e-3][10][-13][plot_data/ver3/MUMIMO16x4_interf_64QAM_TFIRC_1p_4UE/val_16000_CDLuma_velo10_15_TFIRC_mcss11_12_int1_v8_1intfixcsv_data/mod06bler_bin_plot_per_pilot.csv]

\revised{The same UMa trained model was also tested using the UMi channel model, with otherwise similar simulation parameters as above. The results are shown in Fig.~\ref{fig:bler_umi_15}. The performance gain of the proposed EqDeepRx model over the baseline is largely similar to the UMa scenario, even though the model never received any training with the UMi channel model. This indicates that the model has generalized and can perform over a wide range of channel conditions.}

\plotbler{plot_data/rev/ver1/MUMIMO16x4_ch_uma_interf_64QAM_deeprx_standalone_denoise_12p_4UE_pre3x64_dil142_highDS_8gpu_lr4e_05_newint2_WIPsmooth_8cbec91d/val_32000_CDLumi_velo10_15_mcss11_12_int1_v8_1intfixcsv_data/mod06bler_bin_plot_per_pilot.csv}[\revised{BLER with UMi and UE speed of 10--15 m/s. Data rate matched MCSs.}][fig:bler_umi_15][BLER][5e-3][10][-13]

\begin{figure}[!t]
	\centering
	                \includegraphics{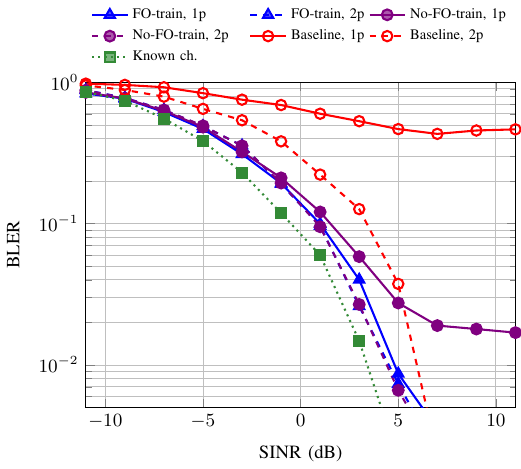} \removewhitespace
	\caption{\revised{BLER with CDL-C, UE speed of 30--35 m/s and FO of $150$ Hz. Data rate matched MCSs. ``FO-train'' and ``No-FO-train'' are both EqDeepRx, differing only in whether FO was present during training.}}
	\label{fig:bler_fo}
\end{figure}

\revised{To gauge EqDeepRx performance under hardware impairments, simulations were also performed with local oscillator (LO) frequency offset (FO) in the highest-speed scenario (30--35~m/s) of Fig.~\ref{fig:cdl_c_0_5ms_and_30_35ms}(b), where channel aging is most pronounced. The LO-induced FO was modeled as a random static offset per slot, drawn from a zero-mean normal distribution with a standard deviation of $150$ Hz. We trained a new model with samples containing varying levels of FO (``FO-train''), and also evaluated our original model, which never saw FO during training (``No-FO-train''), directly on the FO-impaired data; the BLER results are shown in Fig.~\ref{fig:bler_fo}. As in the FO-free case, the FO-train model outperforms the baseline by approximately 2--2.5~dB at 10\% BLER and performs best with 1 pilot, whereas the baseline requires 2 pilots to estimate the FO reliably. We noticed that when validating at lower speeds, the No-FO-train model performs nearly identically to the FO-train model, indicating considerable generalization to this unseen impairment. At this highest speed, however, the No-FO-train model in 1-pilot case exhibits a BLER error floor which is not present in the FO-train and both 2 pilot cases. Thus, explicitly representing FO during training becomes increasingly important for the more challenging single-pilot configuration as UE speed increases.}

\tikzsetnextfilename{}

\revised{The BLER performance of EqDeepRx was also validated with a lower modulation order by reducing the MCSs to 9 and 10 for the 1-pilot and 2-pilot cases, respectively. These correspond to the highest code rates for 16-QAM transmissions in MCS Table 2. In this case, we trained another model to correspond to this scenario and the results are shown in Fig.~\ref{fig:bler_uma_15_16qam}, from where it can be observed that EqDeepRx achieves comparable performance also with the lower modulation order. Indeed, the performance gain over the baseline is nearly identical to the corresponding scenario with 64-QAM transmissions in Fig.~\ref{fig:cdl_c_10_15ms}(b). We have also verified internally that a single, albeit slightly larger, model can achieve the same performance across multiple modulation schemes via per-modulation output masking of bits.}

\plotbler{plot_data/rev/ver1/MUMIMO16x4_ch_uma_interf_16QAM_deeprx_standalone_denoise_12p_4UE_pre3x64_dil142_highDS_8gpu_lr4e_05_newint2/val_16000_CDLC_velo10_15_mcss11_12_int1_v8_1intfix_v2csv_data/mod04bler_bin_plot_per_pilot.csv}[\revised{BLER with CDL-C and UE speed of 10--15 m/s.  MCSs 9--10, corresponding to 16-QAM modulated transmit signal.}][fig:bler_uma_15_16qam][BLER][5e-3][11][-5]

Altogether, these validation results indicate robust performance by the EqDeepRx model. It can provide a substantial performance gain across all the evaluated channel conditions and models. \revised{We} have also done several additional validation experiments, where similar performance was observed, although they have been omitted from this article for brevity.

\subsection{Spectral efficiency}

\begin{figure}[!t]
	\centering
        \includegraphics{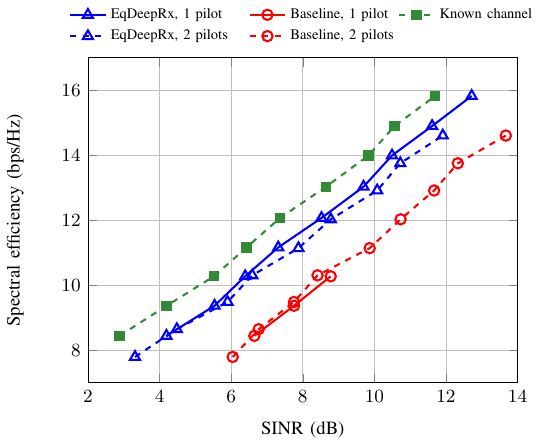}
				\removewhitespace
				\vspace{-0.5mm}
    	\caption{Spectral efficiency, using CDL-C and UE speed of 10--15 m/s.}
	\label{fig:spec_eff}
		\removewhitespace
\end{figure}

To obtain a better understanding of the spectral efficiency improvement achievable by EqDeepRx, additional validations have been performed with a set of MCS values under CDL-C channel model, with UE speeds of 10--15 m/s. The SINR required to achieve a target BLER of 10\% is measured by averaging over multiple channel realizations with each MCS and DMRS configuration. These SINR values are then coupled with the corresponding spectral efficiency achievable with that particular MCS and DMRS configuration. This yields individual data points that can be plotted on an SINR vs. spectral efficiency plot. In particular, the spectral efficiency is defined as follows:
\begin{equation}
\mathrm{SE} = N_T Q_m r (1-\mathrm{BLER}_\mathrm{target}) \rho \gamma, \label{eq:se}
\end{equation}
where $N_T$ is the number of MIMO layers, $Q_m$ is the number of bits per symbol (6 with 64-QAM), $r$ is the code rate, $\mathrm{BLER}_\mathrm{target}$ is the target BLER of 10\%, $\rho$ is the proportion of data-carrying REs ($\nicefrac{13}{14}$ or $\nicefrac{12}{14}$ with 1 and 2 pilots, respectively), and $\gamma$ is the proportion of the useful signal duration in the OFDM symbol ($\gamma \approx 0.93$ in our simulation scenario). Note that when calculating the spectral efficiency, we use the target code rate as the value of $r$ in \eqref{eq:se}, although the true proportion of information bits varies slightly due to rounding and padding. We observed these discrepancies to be within 1.5\% under our simulation configurations, with typical discrepancies well below 1\%. Therefore, the impact on the results is negligible.

The resulting spectral efficiencies are shown in Fig.~\ref{fig:spec_eff}, where the markers denote the specific SINR points in which the BLER target is achieved, while the line is the result of interpolation between these individual data points. If target BLER cannot be achieved with a certain MCS value, it is omitted from the figure. It can be observed that the performance of the EqDeepRx is consistently high across the whole SINR range, which covers the 64-QAM MCSs from index 11 through 19. Moreover, with the higher MCSs, it seems to be beneficial to use the 1 pilot configuration with EqDeepRx since it provides the highest spectral efficiency. The baseline is not able to benefit from the 1 pilot configuration since it requires 2 pilots in order to achieve the target BLER with MCSs above 13. Altogether, the spectral efficiency gain of EqDeepRx over the baseline solution is approximately 15--25\%, depending on the SINR level. This is essentially the throughput gain achievable by the proposed scheme.

\subsection{Model complexity and ablation studies}

\begin{figure}[!t]
    \centering
        \includegraphics{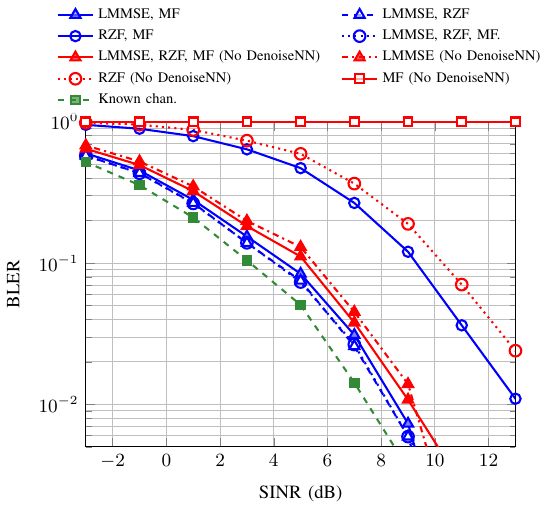}
				\removewhitespace
    \caption{Ablation study on the parallel equalizers in 1-pilot setup. CDL-C, speed 10--15 m/s.}
    \label{fig:ablation_eq}
		\removewhitespace
\end{figure}

A key design decision for the developed EqDeepRx algorithm is the selection of the equalizer(s) that provides the soft symbol estimates for the subsequent ML processing. To understand better the impact of different equalization approaches on the final model performance, Fig.~\ref{fig:ablation_eq} shows the BLER for CDL-C validation under four different equalizer combinations. In addition to the adopted structure with parallel LMMSE and RZF equalizers, we also evaluated the performance with some combinations that included a matched filter (MF) type approach. \revised{This equalizer consists of a matched filter in spatial domain, which corresponds to an output of 
$\hat{\mathbf{x}}_{ij,\mathrm{MF}} = \mathbf{D}_{ij,\mathrm{MF}}^{-1} \widehat{\mathbf{H}}_{ij}^\hop \mathbf{y}_{ij}$, where $\mathbf{D}_{ij,\mathrm{MF}}=\diag(\widehat{\mathbf{H}}_{ij}^\hop \widehat{\mathbf{H}}_{ij})$. We omitted the normalization in the simulations and used $\mathbf{D}_{ij,\mathrm{MF}}=\mathbf{I}$, which did not have an effect on the results.}

From Fig.~\ref{fig:ablation_eq} it is evident that the combinations incorporating the LMMSE equalizer perform the best in terms of BLER. The option with LMMSE and RZF equalizers, corresponding to the primary EqDeepRx architecture, achieves the lowest BLER, while adding the MF equalizer has no impact on the performance. Utilizing LMMSE and MF equalizers has a slightly worse performance. Together, these observations indicate that the information provided by the MF is insufficient compared to the RZF output from the perspective of the subsequent ML processing. If removing the LMMSE equalizer and using only RZF and MF equalizers, the performance drops by approximately 5~dB due to inter-cell interference.

Additionally, we note that we could not train a functional model by utilizing solely the LMMSE equalizer with DenoiseNN included in the model. Our hypothesis for the instability is that, when DenoiseNN pre-processes input data with initial random weights, the LMMSE calculation with INCM estimation becomes numerically too unstable to proceed with optimization. When another EQ such as RZF is included, this alternative EQ can pass sufficient information for training of neural network components even at the initial stage when the neural network weights are near random. Figure~\ref{fig:ablation_eq} also includes options without DenoiseNN in which case also an LMMSE-only model can be trained. Nevertheless, the model employing both LMMSE and RZF is performing best also among models without DenoiseNN, although falling short of the corresponding model that incorporates DenoiseNN.

\begin{figure}[!t]
	\centering
        \includegraphics{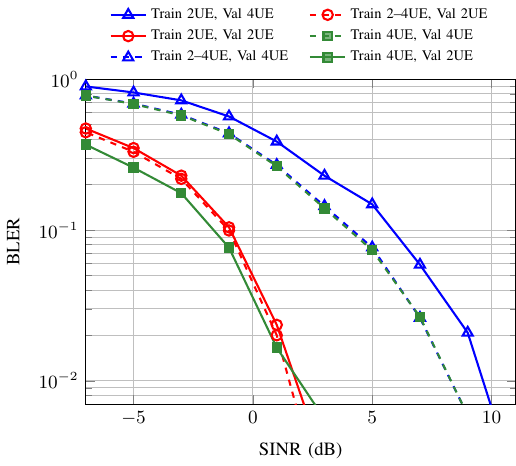}
   	\caption{\revised{Generalization across MIMO layer counts (two versus four UEs) using a one-pilot configuration. The model trained with four layers generalizes to two layers without a performance loss, whereas the reverse mismatch causes a degradation. CDL-C channel with UE speeds of 10--15 m/s.}}
	\label{fig:UEcount3}
	\removewhitespace
\end{figure}

Next, let us evaluate EqDeepRx's capability of dealing with an arbitrary number of MIMO layers. This feature was tested by evaluating different EqDeepRx models with 2 and 4 layers. In particular, three different EqDeepRx models were trained with data consisting of 2 UEs, 4 UEs, and 2--4 UEs. Similar to the primary simulation assumptions, each UE has one transmit antenna, which means that the number of UEs directly translates to the number of overlapping MIMO layers.

\revised{The corresponding BLER results are shown in Fig.~\ref{fig:UEcount3}. Firstly, it is evident from the results that the EqDeepRx model generalizes well to different numbers of MIMO layers when the training data includes examples of all options. This is evidenced by the fact that the model trained with 2--4 MIMO layers achieves identical performance as the models that have been trained with the prevailing number of layers, in both cases of 2 and 4 UEs. More importantly, the model trained only with 4 UEs generalizes to 2 UEs without a performance loss and is even slightly better in the low-SNR region than the model trained with 2 UEs. This slightly better performance may be occurring because four-layer training provides twice as many layer-wise training examples. In addition, although the inter-cell interference conditions are the same for both layer counts, the small difference may be related to their different effective operating points and to the larger residual errors that may remain after equalization in the four-layer case, thus providing more challenging training examples. In the opposite direction, the model trained only with 2 UEs falls short of the generalized model when evaluated with 4 UEs, with a performance deficit of approximately 1.5~dB. These results support layer-count independence when reducing the number of layers, while previously unseen higher layer counts remain more challenging and should be represented in the training data.}

\begin{figure}[!t]
	\centering
        \includegraphics{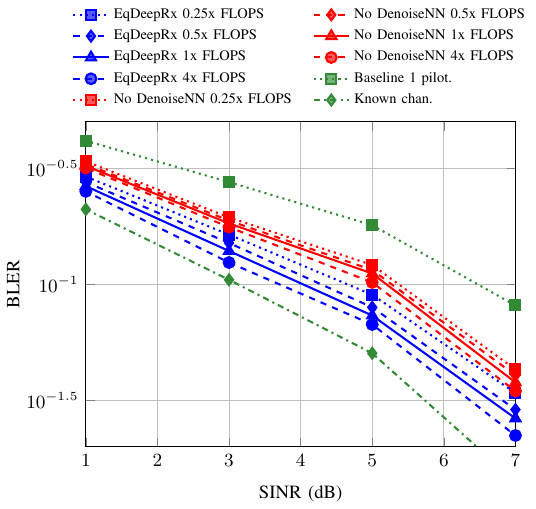}
				\removewhitespace
    	\caption{Models with different FLOPs complexity in 1-pilot setup. CDL-C, speed 10--15 m/s.}
	\label{fig:FLOPS1}
	\removewhitespace
\end{figure}

Then, to investigate the effect of model complexity on the BLER performance, we trained EqDeepRx models of different sizes and architectures and evaluated them with CDL-C channel, focusing on the 1-DMRS scenario. Two different architectures were considered: the full EqDeepRx model and a reduced EqDeepRx model without the DenoiseNN block. Both architectures were trained with four different model sizes, which are quantified in terms of number of FLOPs with respect to the primary EqDeepRx model architecture.

The results of this experiment are collected in Fig.~\ref{fig:FLOPS1}. Firstly, it can be observed that none of the architectures without the DenoiseNN block can outperform the full EqDeepRx model, regardless of complexity. Even the model with four times higher complexity in the detector and demapper CNNs falls 0.2~dB short of the smallest EqDeepRx model that has 16 times fewer FLOPs. This indicates that it is crucial to allocate the ML processing to the right parts of the receiver, since that allows for achieving the best balance between complexity and performance. As for the EqDeepRx models, allocating more FLOPs to the model does seem to yield consistently higher performance. Quadrupling the FLOPs of the primary EqDeepRx architecture provides a gain of 0.2--0.3~dB in terms of BLER, at least in this particular scenario. However, we wish to emphasize that the exact amount of gain most likely depends on the adopted simulation scenario, especially the MCS. Nevertheless, these observations demonstrate that the proposed EqDeepRx receiver can be flexibly optimized for the desired performance and complexity targets, which depend on the available hardware, power, and throughput requirements. Such flexible optimization might not be possible with conventional non-ML receivers.

\begin{figure}[!t]
	\centering
        \includegraphics{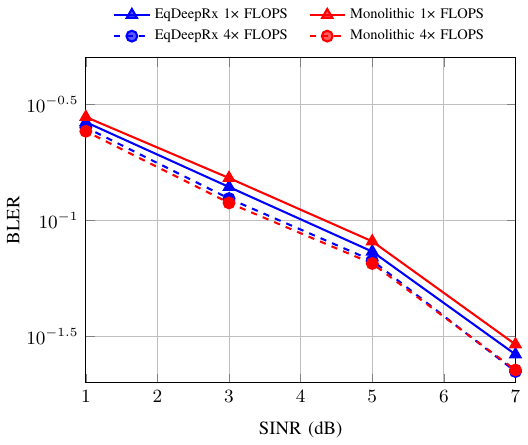}
				\removewhitespace
    	\caption{EqDeepRx vs Monolithic detector and demapper. All options include DenoiseNN and LMMSE and RZF equalization. CDL-C, speed 10–15 m/s.}
	\label{fig:monolithic_comparison}
	\removewhitespace
\end{figure}

\revised{We also evaluated a monolithic alternative that replaces only EqDeepRx's per-layer detector and demapper. It concatenates the $N_T=4$ layers along the feature axis and processes them jointly (unlike EqDeepRx's DetectorNN, which processes each MIMO layer independently). Its detector input additionally includes the output of a matched filter (MF) equalizer alongside the LMMSE and RZF equalizer outputs used by EqDeepRx; these three complex-valued equalizer outputs and two real-valued index maps give eight real channels per layer, hence an input shape of $[F,S,8N_T]=[192,14,32]$. Unlike EqDeepRx, its pointwise convolutions are not grouped by layer and therefore can mix information across all layers. Its $N_TB=32$ outputs are reshaped to $[N_T,F,S,B]=[4,192,14,8]$. This joint processing ties the network dimensions to $N_T$, whereas EqDeepRx's shared per-layer networks support different layer counts without architectural changes. Apart from the cross-layer mixing, the detector and demapper use the subsampled architecture described above.}

\revised{We selected the widths and depths of the four monolithic variants in Table~\ref{tab:compute_plus_total_params} to match the per-layer complexity of EqDeepRx; at the $1\times$ and $4\times$ points, the detector uses 6 and 8 residual blocks with per-MIMO-layer widths of 32 and 76, and DenoiseNN widths of 48 and 64. Dividing each joint forward-pass cost by $N_T$ gives 0.36 and 1.48 GFLOPs per layer, versus 0.37 and 1.44 GFLOPs for EqDeepRx; the monolithic models contain 346k and 1751k parameters, with no monolithic-specific hyperparameter tuning.}

\revised{As shown in Fig.~\ref{fig:monolithic_comparison}, EqDeepRx slightly outperforms the monolithic model at the smaller operating point, while the two are nearly identical at the larger operating point. Thus, the layer-independent EqDeepRx design incurs no observable performance penalty relative to the FLOPs-matched monolithic alternative.}

\begin{table}[t]
	\centering
	\scriptsize
	\setlength{\tabcolsep}{4pt}
	\begin{tabular}{lcccccc}
		\hline

		Model & \multicolumn{4}{c}{GFLOPs} & \shortstack{Params\\(k)} & \shortstack{\\SINR@\\10\%BLER} \\
		\cline{2-5}
		& Denoise & Detect & Demap & Total & & \\
		\hline

		EqDeepRx 0.25x       & 0.02 & 0.06 & 0.02 & 0.10 &   29 & 4.73 \\

		EqDeepRx 0.5x        & 0.03 & 0.10 & 0.05 & 0.18 &   55 & 4.43 \\

		EqDeepRx 1x          & 0.05 & 0.26 & 0.06 & 0.37 &  116 & 4.19 \\

		EqDeepRx 4x          & 0.16 & 1.07 & 0.21 & 1.44 &  472 & 3.85 \\

		NoDenoise 0.25x      & --   & 0.08 & 0.02 & 0.10 &   26 & 5.54 \\

		NoDenoise 0.5x       & --   & 0.14 & 0.05 & 0.19 &   49 & 5.41 \\

		NoDenoise 1x         & --   & 0.30 & 0.06 & 0.37 &  104 & 5.31 \\

		NoDenoise 4x         & --   & 1.21 & 0.24 & 1.45 &  427 & 5.06 \\

		Monolithic 0.25x & 0.01 & 0.05 & 0.04 & 0.10 & 96 & 5.22\\

		Monolithic 0.5x & 0.02 & 0.10 & 0.05 & 0.18 & 176 & 4.72\\

		Monolithic 1x        & 0.03 & 0.18 & 0.15 & 0.36 &  346 & 4.47 \\

		Monolithic 4x        & 0.05 & 1.20 & 0.23 & 1.48 & 1751 & 3.70 \\

		No LMMSE 1x            & 0.05 & 0.26 & 0.06 & 0.37 &  116 & 9.48 \\
		No LMMSE 4x          & 0.16 & 1.07 & 0.21 & 1.44 &  472 & 9.35 \\
		\hline
	\end{tabular}
	\caption{Per-part and total GFLOPs per inference per 16 PRB (192 subcarriers) / 1 MIMO layer. Also parameter counts (equalizers excluded) and performance are shown. }
	\label{tab:compute_plus_total_params}
	\removewhitespace
\end{table}

\begin{figure}[!t]
	\centering
        \includegraphics{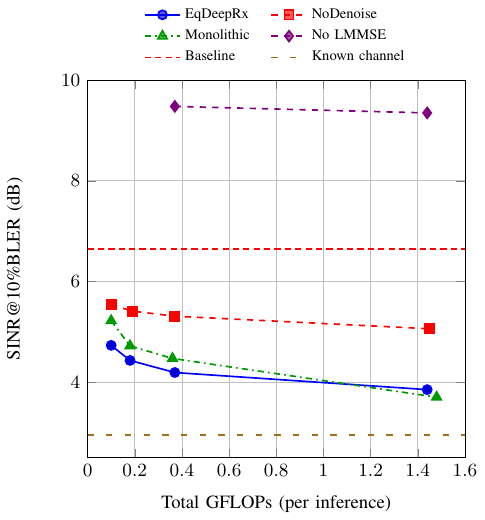}
    	\caption{SINR vs.\ total compute across DeepRx model families and hyperparameter variants. Baseline is the performance of the non-learned LMMSE baseline. Each dot represents a trained model with different hyperparameters. CDL-C, speed 10–15 m/s.}
	\label{fig:gfops_vs_sinr}
	\removewhitespace
\end{figure}

The findings and details regarding the complexity and performance of all the different ML receiver model variants are collected in Table~\ref{tab:compute_plus_total_params}. It lists the FLOPs per inference for each ML block, as well as the number of trainable parameters. Note that the non-ML receiver elements are not considered in the table, since the focus is on the additional complexity of the ML processing. \revised{The number of FLOPs reported for each EqDeepRx variant can be considered to represent the additional compute required over the baseline, in order to achieve the associated performance gain.} Table~\ref{tab:compute_plus_total_params} also shows the SINR required to achieve the target BLER of 10\% in the considered simulation scenario (CDL-C, UE speed of 10--15 m/s, 1 DMRS, MCS 11). The primary EqDeepRx architecture has an additional complexity of 0.37 GFLOPs per inference, with most of the complexity residing in the DetectorNN block. The most complex architecture evaluated is the monolithic 4x model with a complexity of 1.48 GFLOPs per inference, and it is able to outperform the primary EqDeepRx approach by only 0.5 dB. This indicates the proposed EqDeepRx model demonstrates a rather favorable balance between performance and complexity. \revised{It should be noted that the FLOP counts reported herein should be interpreted solely as theoretical indicators of computational complexity, while corresponding energy consumption and inference latency depend on the underlying hardware implementation. Detailed assessments of energy efficiency and latency are better suited to hardware-oriented studies.}

To further evaluate the performance--complexity trade-off of the considered augmented receiver models, Fig.~\ref{fig:gfops_vs_sinr} visualizes the SINR and FLOPs results of Table~\ref{tab:compute_plus_total_params}. It can be observed that the proposed EqDeepRx model achieves the highest performance in the lower FLOPs regime, while the monolithic approach slightly outperforms it under higher complexity. Moreover, it can be seen that the effect of the additional DenoiseNN is approximately 1 dB, regardless of the model complexity. The model variant without LMMSE equalizer falls clearly short of all other options, regardless of the number of FLOPs. Altogether, the shape of the curves indicates that additional complexity yields diminishing returns when the model size is increased. With EqDeepRx, we have aimed at operating in the regime where the model complexity has a clear impact on the receiver performance.

\begin{figure}[!t]
	\centering
	\includegraphics{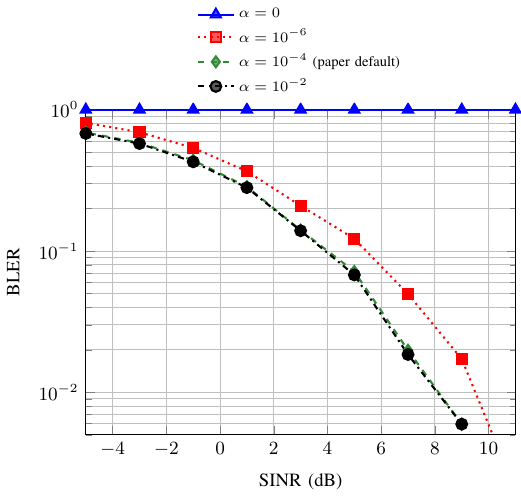}
				\removewhitespace
    	\caption{\revised{BLER on CDL-C, 10--15\,m/s, 1 pilot, with interference.
	         Sensitivity to the RZF regularizing term $\alpha$.}}
	\label{fig:lmmsealpha_sensitivity}
\end{figure}
\revised{Finally, we performed a brief sensitivity analysis with respect to selected hyperparameters, namely the symbol-loss regularization weight $\lambda$, the RZF regularizing term $\alpha$, the learning rate $\eta$, and the interference coherence bandwidth. Overall, EqDeepRx was observed to be robust to the exact choice of these design parameters, with the exception of the learning rate. We evaluated the impact of the symbol-loss regularization weight $\lambda$ by sweeping it between $0$, $10^{-5}$ (paper default), $10^{-4}$, and $10^{-1}$, and found that, in the considered evaluation scenario, $\lambda$ has essentially no impact on the final BLER performance. This is most likely due to the specific nature of the evaluation scenario, and we expect that with certain channel conditions and/or model hyperparametrizations, having $\lambda > 0$ can help with model convergence and final performance. Indeed, this is what we observed during model development, which motivated including such a term in the loss function. Moreover, it ensures that the output of the DetectorNN represents actual soft symbol estimates, instead of some latent variable.}

\revised{The sensitivity to the RZF regularizing term $\alpha$ is shown in Fig.~\ref{fig:lmmsealpha_sensitivity}, where $\alpha$ was swept between $0$, $10^{-6}$, $10^{-4}$ (paper default), and $10^{-2}$. Except for the very small values $\alpha=0$ and $\alpha=10^{-6}$, the BLER performance is virtually unaffected by this choice, confirming that EqDeepRx is not sensitive to the exact value of the RZF regularizing term, as long as it is not chosen too close to zero.} 

\revised{Similarly, we evaluated the sensitivity to the interference coherence bandwidth, i.e., the interval of subcarriers over which the interference-plus-noise covariance matrix is assumed constant and thus estimated jointly, comparing the paper default of 24 subcarriers against a narrower interval of 12 and a wider interval of 48 subcarriers. The BLER performance was found to be practically unaffected by this choice within the evaluated range.}

\revised{As for the learning rate, its impact on the model performance is quite significant, as expected. We observed typical sensitivity where significantly increasing or decreasing the learning rate from the selected value of $4.4 \times 10^{-3}$ leads to a clear increase in BLER, with the chosen value providing the highest performance among the evaluated ones.}

\section{Conclusion}
\label{sec:conc}

This article presented \emph{EqDeepRx}, a practical ML-aided MIMO OFDM receiver that combines expert signal processing with compact neural components to deliver high link performance at low complexity. The design (i) denoises DMRS-based channel estimates with a lightweight DenoiseNN, (ii) runs two complementary linear equalizers (LMMSE and RZF) in parallel to avoid learning matrix inverses, and (iii) performs per-MIMO-layer data-aided detection and demapping with shared weights, enabling near-linear complexity growth with layer count and layer-count invariance. Building on earlier DeepRx insights, all variants observe the full time-frequency OFDM grid and allow DetectorNN to learn a data-aided detection scheme.

\revised{Extensive simulations with 5G/6G-compliant OFDM waveforms and LDPC channel coding show consistent improvements over conventional and ML baselines. Across CDL-C, CDL-D, UMi and UMa channel models, \emph{EqDeepRx} reduces the SINR at 10\% BLER by about $2$~dB (CDL-C) and by more than $4$~dB (UMi/UMa), and improves spectral efficiency by roughly $15$–$25\%$.} It maintains comparable BLER with one or two DMRS symbols and remains robust up to 30–35~m/s; in contrast, the baseline struggles with one DMRS under high mobility.

Ablation studies show that (a) the LMMSE+RZF equalizer pair is essential, adding MF brings no benefit, and removing interference-handling LMMSE degrades performance by up to 5 dB; (b) DenoiseNN is critical; omitting it costs 1~dB even with a larger detector/demapper; and (c) per-layer detection/demapping matches or exceeds monolithic designs while uniquely supporting arbitrary MIMO layer counts.

In summary, EqDeepRx delivers robust BLER and throughput gains over a conventional receiver while preserving low computational complexity and deployment practicality. By keeping classical estimation and equalization in the loop and focusing learning capacity where it has the highest payoff (channel denoising and data-aided detection), EqDeepRx avoids emulating matrix inverses and remains resilient under interference, mobility, and channel variability.

As future work, we will investigate reducing DMRS density and handling RF-chain nonlinearities (e.g., PA saturation, IQ imbalance, quantization) within \emph{EqDeepRx}.

\bibliographystyle{IEEEtran}

\end{document}